\documentclass[%
 reprint,
superscriptaddress,
 amsmath,amssymb,
 aps,
pra,
]{revtex4-2}

\usepackage[english]{babel}
\usepackage{comment}
\usepackage{tikz}
\definecolor{fruchtblue}{RGB}{0,0,0}
\usepackage{graphicx}
\usepackage{dcolumn}
\usepackage{bm}

\usepackage{tikz}
\usepackage{pgf}
\usetikzlibrary{arrows, automata}
\usepackage{hyperref}
\usetikzlibrary{fit,positioning}
\hypersetup{
    colorlinks=true,
    linkcolor=black,
    urlcolor=black,
    citecolor=black
}
\usepackage{physics}
\usepackage{quantikz}
\usetikzlibrary{angles,quotes}

\usepackage{comment}

\begin{document}


\title{Entanglement Dynamics in Katz-Weighted Graph States}

\author{Lucio De Simone}
\email{l.desimone3@student.unisi.it}
\affiliation{DSFTA, University of Siena, Via Roma 56, 53100 Siena, Italy}
\affiliation{INFN Sezione di Perugia, 06123 Perugia, Italy}
\author{Lorenzo Capra}
\affiliation{DSFTA, University of Siena, Via Roma 56, 53100 Siena, Italy}
\affiliation{INFN Sezione di Perugia, 06123 Perugia, Italy}
\author{Roberto Franzosi}
\affiliation{DSFTA, University of Siena, Via Roma 56, 53100 Siena, Italy}

\affiliation{INFN Sezione di Perugia, 06123 Perugia, Italy}

\date{\today}

\begin{abstract}
We investigate the entanglement dynamics of quantum states defined on graphs with non-local Ising interactions governed by the Katz kernel of the underlying network. The interaction pattern is physically motivated by a gapped fermionic mediator propagating on the same graph, whose perturbative elimination yields an effective Katz-weighted Ising Hamiltonian. Using the Entanglement Distance, we derive an exact analytical expression for the entanglement generated from an initially separable state and apply it to representative deterministic graph families.

We then characterize the dynamics in different propagation regimes. In the weak-Katz regime, the dynamics admits a systematic motif expansion with triangles entering at first order order and four-cycles, local degree structure, and overlappin triangles appearing at second order. In the strong-propagation regime, the interaction is instead dominated by the principal adjacency mode and by the localization properties of its eigenvector. For Erd\H{o}s--R\'enyi graphs, the weak-propagation expansion can be averaged analytically, revealing a
locally tree-like contribution in the sparse regime and saturation of the Entanglement Distance density in the dense regime.

Our results connect entanglement dynamics with both the walk-based and spectral structure of complex networks.

\end{abstract}

\maketitle

\section{Introduction}

Graph states provide a natural bridge between graph theory and multipartite quantum entanglement. In the standard stabilizer formalism, qubits are assigned to the vertices of a graph and entanglement is generated by controlled-phase gates acting along its edges. A natural generalization consists of allowing pair-dependent interaction strengths, leading to weighted graph states. These states arise as the dynamical output of Ising-type Hamiltonians with non-uniform couplings and offer a flexible framework for investigating many-body entanglement and measurement-based protocols \cite{PhysRevLett.86.910, PhysRevLett.86.5188, PhysRevA.69.062311, PhysRevLett.97.150504, Hartmann_2007,Anders_2007}. They also appear naturally in quantum-network settings, where multipartite entangled states have to be generated and distributed over a prescribed connectivity graph \cite{Perseguers_2013,Wallnofer2019-zs,Hahn2019-kw, PhysRevA.108.062614}.

A different but related perspective comes from network theory, where matrix functions of the adjacency matrix are used to assign weights to walks on a graph. Katz centrality was introduced as a walk-based measure in which longer walks are suppressed by an attenuation factor \cite{Katz_1953}. More generally, resolvent and matrix-function methods are used to define network communicability, where the relation between two nodes receives contributions from walks of all lengths rather than only from shortest paths \cite{PhysRevE.77.036111, 10.1093/comnet/cnt007, ESTRADA201289}. The corresponding Katz walk kernel is
\begin{equation}
K=A(I-\beta A)^{-1}
=
A+\beta A^2+\beta^2A^3+\cdots \, .
\label{eq:Kdef}
\end{equation}
Here, the element $(A^k)_{ij}$ counts the number of walks of lenght $k$ from $i$ to $j$. Thus $K_{ij}$ counts all walks from $i$ to $j$, with longer walks suppressed by powers of $\beta$.

In this work we use the Katz kernel as the coupling matrix of a weighted graph state. In this way, qubits are coupled not only through direct edges, but also through indirect walks on the graph. Katz-weighted graph states therefore provide a natural setting in which to ask how local connectivity and longer-range graph structure affect multipartite entanglement. Related graph-based questions also arise in ensemble-based networks, including entanglement percolation, random graph states, and graph-based random states \cite{PhysRevLett.103.240503,Collins_2013,PhysRevA.89.052335,52xz-3hpc}. To quantify the generated multipartite entanglement, we use the Entanglement Distance (ED), a geometric entanglement measure derived from the Fubini--Study metric on projective Hilbert space initially introduced in \cite{cocchiarella_entanglement_2020} and summarized in \cite{e28030299} and applied to various systems\cite{vafafard_multipartite_2022,nourmandipour_entanglement_2021, Vesperini_2023, vesperini_correlations_2023}. Recently, it has been applied to states defined on graph structures \cite{DeSimone_2025,https://doi.org/10.1002/qute.202500514}. This measure has also been extended to multipartite mixed states \cite{vesperini_entanglement_2023} and present an extension for applying to continuous variable systems \cite{Vesperini_2024}.

The paper is organized as follows. In Sec.~\ref{sec:model} we derive the effective Katz-weighted Ising Hamiltonian from a simple fermionic model. In Sec.~\ref{sec:entanglement} we introduce the Entanglement Distance, derive its exact expression under the Katz-weighted dynamics, and apply it to representative deterministic graphs. In Sec.~\ref{sec:propagation}
we characterize the entanglement dynamics across different propagation regimes. We first develop the weak-propagation expansion, which relates
the first corrections to local graph structures, and then analyze the strong-propagation regime close to the Katz threshold. Finally, we extend the weak-propagation analysis to random graph ensembles and obtain
analytical results for sparse and dense Erd\H{o}s--R\'enyi graphs.

\section{Katz-weighted Ising Hamiltonian}
\label{sec:model}
Let $G=(V,E)$ be a simple undirected graph with vertex set $V=\{1,\dots,M\}$, edge set $E$ with $L=|E|$, and adjacency matrix $A$. A spinless fermionic mode, with annihilation (creation) operator $c_i$ ($c^\dagger_i$), is associated with each vertex. The quadratic Hamiltonian of the fermionic mediator is
\begin{equation}
    H=\sum_{i,j}c_i^\dagger Q_{ij}c_j-\frac12\operatorname{Tr} Q\, ,
    \qquad
    Q=\mu I-w A \, .
    \label{eq:Qdef}
\end{equation}
Here $\mu$ is the detuning of the mediator, while $w\ge0$ is the hopping amplitude along graph edges. We work in the gapped regime in which $Q$ is positive definite. Since the eigenvalues of $Q$ are $q_\alpha=\mu-w\lambda_\alpha(A)$, this is ensured by $\mu>w\lambda_1$, where $\lambda_1$ is the largest eigenvalue of $A$. Physically, this condition means that the hopping can lower the mediator energies by at most $w\lambda_1$, while the lowest mediator excitation remains separated from the vacuum by the positive gap $\mu-w\lambda_1$.  
It is useful to rewrite the mediator in terms of Majorana operators,
$a_i=c_i+c_i^\dagger$ and $b_i=-i(c_i-c_i^\dagger)$. Since $A$ is symmetric and $\mu,w$ are real, the matrix $Q$ is symmetric. Then, the quadratic Hamiltonian becomes
\begin{equation}
    H=\frac{i}{2}\sum_{i,j}a_iQ_{ij}b_j \, .
\end{equation}
Since the qubit operators are fermion-parity even, the interaction Hamiltonian must also be even under fermion parity. A term proportional to $a_i\sigma_z^{(i)}$ would instead be parity odd, because a single Majorana operator changes the fermion parity. We therefore introduce an auxiliary Majorana operator $\chi$ and use the parity-even coupling
\begin{equation}
    V=ig\chi\sum_i a_i \sigma_z^{i} \, .
    \label{eq:Vcoupling}
\end{equation}
Here $g$ is some coupling constant and $\chi$ is an auxiliary Majorana mode external to the graph. It may be viewed as a Klein factor associated with an additional fermionic sector, in the sense that Klein factors can be represented as additional Majorana degrees of freedom coupled to physical Majoranas \cite{PhysRevLett.110.216803}. Since $\chi$ is not attached to any vertex, it carries no graph index and does not modify the graph structure encoded in $Q$. Diagonalizing $Q$ and eliminating the gapped fermionic modes to second order in $g$ by a Schrieffer--Wolff transformation gives
\begin{equation}
\label{hamil}
    H_{\rm eff}=-2g^2\sum_{i<j}(Q^{-1})_{ij}\sigma_z^{i} \sigma_z^{j}-g^2\sum_i (Q^{-1})_{ii}I \, .
\end{equation}
We introduce the dimensionless parameter $
\beta=w/\mu$. Then, the positivity condition on $Q$ becomes $0\le\beta<\beta_c$, where $\beta_c=1/\lambda_1$. Since the entangling coupling between pairs of qubits involves off-diagonal parts of $Q^{-1}$, one obtains, for $i\neq j$,
\begin{equation}
    (Q^{-1})_{ij}=\frac{\beta}{\mu}\,[A(I-\beta A)^{-1}]_{ij} \, .
    \label{eq:QtoK}
\end{equation}
Thus, the effective Hamiltonian, up to additive constants, takes the form, relabeling,
\begin{equation}
\label{hamiltoniana}
H=-\frac{J}{2}
\sum_{i< j}
K_{ij}
\sigma_z^{i}
\sigma_z^{j} \, ,
\end{equation}
where $J=4g^2\beta/\mu$ and $K$ is given in Eq. \eqref{eq:Kdef}. Thus the Schrieffer--Wolff transformation generates an Ising Hamiltonian whose off-diagonal couplings are Katz-weighted. The same condition that keeps the mediator gapped also guarantees the convergence of the Katz expansion, while $\beta$ controls the relative weight of walks of increasing length.

\section{Entanglement Distance and Dynamics}
\label{sec:entanglement}
In this paper, we use the Entanglement Distance (ED), an entanglement quantifier whose geometric formulation is based on the Fubini--Study metric on projective Hilbert space \cite{e28030299}. For a pure state $|\psi\rangle$ of $M$ qubits, it is defined as
\begin{equation}
\label{entanglement}
E(|\psi\rangle)=\sum_{i=1}^M E_i\, ,
\quad
E_i
=
1-
\sum_{a=x,y,z}
\langle \psi|\sigma_a^{i}|\psi\rangle^2  \, ,
\end{equation}
where $\sigma_a^{i}$, with $a=x,y,z$, is the Pauli matrix acting on qubit $i$.  Each contribution $E_i$ has a direct one-versus-rest interpretation.
Denoting by $\bar i=V\setminus\{i\}$ the remaining qubits and by
$\rho_i=\operatorname{Tr}_{\bar i}|\psi\rangle\langle\psi|$ the reduced
state of qubit $i$, its Bloch representation gives
$E_i=2[1-\operatorname{Tr}(\rho_i^2)]$. Thus $E_i$ is the normalized
linear entropy of the reduced qubit and quantifies the entanglement
across the bipartition $i|\bar i$. For a pure global state, $E_i=0$
when qubit $i$ is unentangled from the rest, while $E_i=1$ when
$\rho_i=I/2$. Consequently, the total ED vanishes for fully separable
pure states and reaches its maximal value $M$ when all single-qubit
reduced states are maximally mixed.

We consider the unitary evolution generated by the Hamiltonian in Eq.~\eqref{hamiltoniana}, starting from the initial product state $|+\rangle^{\otimes M}$, where
$|+\rangle=(|0\rangle+|1\rangle)/\sqrt{2}$ and
$|0\rangle$ and $|1\rangle$ are the eigenstates of $\sigma_z$ with eigenvalues $+1$ and $-1$ respectively. For this initial state, the contribution of vertex $i$ to the ED takes the compact form
\begin{equation}
\label{eq:exact_Ei}
E_i(t,\beta)
=
1-
\prod_{j\neq i}
\cos^2\left(tK_{ij}\right)\, ,
\end{equation}
where $t=J\tau$ is the dimensionless rescaled time and $\tau$ is the physical time. Unless stated otherwise, time is measured in terms of the dimensionless variable $t$. The total Entanglement Distance is therefore
\begin{equation}
    E(t,\beta)
    =
    \sum_{i=1}^{M}E_i(t,\beta)\, .
    \label{exact}
\end{equation}
Equation~\eqref{eq:exact_Ei} is exact within the effective Ising model.
The Katz couplings $K_{ij}$ set the characteristic entangling
frequencies entering the dynamics of each vertex. When graph symmetries
make several couplings equivalent, the corresponding factors in
Eq.~\eqref{eq:exact_Ei} combine into powers and the dynamics reduces to
a small number of distinct scales. For a fixed vertex $i$,
Eq.~\eqref{eq:exact_Ei} depends only on the couplings connecting $i$
to the remaining vertices. Vertices with equivalent coupling profiles
therefore have identical single-site entanglement dynamics.

A useful way to resolve this coupling pattern is according to the graph
distance from $i$. Let
$\mathcal S_r(i)=\{j\in V:d(i,j)=r\}$ denote the set of vertices at
distance $r$ from $i$, where $d(i,j)$ is the minimum distance on all the paths connecting $i$ with $j$. Equation~\eqref{eq:exact_Ei} can then be
reorganized as
\begin{equation}
    E_i(t,\beta)
    =
    1-
    \prod_{r\geq1}
    \prod_{j\in\mathcal S_r(i)}
    \cos^2\!\left(tK_{ij}\right) \, .
    \label{eq:distance_shell_ED}
\end{equation}
This representation is particularly natural for the Katz interaction,
since the graph distance fixes the lowest order in $\beta$ at which their Katz coupling appears. Indeed, if $d(i,j)=r$, then
$(A^n)_{ij}=0$ for every $n<r$, and therefore
\begin{equation}
    K_{ij}
    =
    \beta^{r-1}(A^r)_{ij}
    +
    \beta^r(A^{r+1})_{ij}
    +\cdots \, ,
    \label{eq:katz_distance_order}
\end{equation}
Thus, vertices at distance $r$ first enter the Katz interaction at order
$\beta^{r-1}$. Since $(A^r)_{ij}$ counts the number of shortest paths of length $r$
connecting $i$ and $j$, the graph distance determines the first order
at which the coupling appears, while the number of shortest paths
determines the coefficient at that order.

The Katz interaction therefore generates a hierarchy of couplings
around each vertex. Direct neighbors, with $r=1$, contribute already at zeroth order,
vertices at distance two ($r=2$) first appear at order $\beta$, vertices at
distance three at order $\beta^2$, and so on. This hierarchy concerns
the couplings themselves. Its effect on the entanglement need not occur
at the same order, since Eq.~\eqref{eq:distance_shell_ED} depends
nonlinearly on each $K_{ij}$.
When all vertices belonging to the same distance shell experience the
same Katz coupling, $K_{ij}=\kappa_i^{(r)}$ for
$j\in\mathcal S_r(i)$, Eq.~\eqref{eq:distance_shell_ED} further
reduces to
\begin{equation}
   \!\!\!\! E_i(t,\beta)
    =
    1-
    \prod_{r\geq1}
    \cos^{2n_i(r)}
    \!\left(t\kappa_i^{(r)}\right) \, ,
    \
    n_i(r)=|\mathcal S_r(i)| \, .
    \label{eq:shell_symmetric_ED}
\end{equation}

The distance-shell representation describes how the Katz interaction
extends through successive walks from a given vertex. A complementary
description is obtained in the spectral basis of the graph. Since $K$
is a matrix function of the adjacency matrix, it has the same
eigenvectors as $A$. Writing
\begin{equation}
A=
\sum_{\alpha=1}^M
\lambda_\alpha
u_\alpha u_\alpha^T \, ,
\end{equation}
with orthonormal eigenvectors $u_\alpha$, the Katz kernel becomes
\begin{equation}
\label{eq:spectral_katz}
K=
\sum_{\alpha=1}^M
k_{\alpha}
u_\alpha u_\alpha^T \, ,
\quad
k_{\alpha}
=
\frac{\lambda_\alpha}{1-\beta\lambda_\alpha} \, .
\end{equation}
This expression shows explicitly how the Katz couplings are determined by both the spectrum and the eigenvectors of the underlying graph.
While the distance representation organizes the interaction according
to the length and multiplicity of walks, the spectral representation
shows how the same propagation process reweights the spectral modes
of the graph. Indeed, the Ising Hamiltonian in Eq.~\eqref{hamiltoniana}, up to additive energy shifts, can be written as
\begin{equation}
\label{twisting}
H = -\frac{J}{4}\sum_{\alpha=1}^{M} k_\alpha S_{z,\alpha}^2,\quad S_{z,\alpha}=
\sum_{i=1}^{M}u_{\alpha,i}\sigma_z^{i} \, .
\end{equation}
Equation~\eqref{twisting} provides a spectral decomposition of the interaction into quadratic collective-mode contributions. For the generalized operators $S_{\mu,\alpha}=\sum_{i=1}^M u_{\alpha,i}\sigma_{\mu}^i$, with $\mu=x,y,z$, one finds $[S_{\mu,\alpha},S_{\nu,\gamma}]=2i\varepsilon_{\mu\nu\rho}\sum_{i=1}^Mu_{\alpha,i}u_{\gamma,i}\sigma_{\rho}^i$, where $\varepsilon_{\mu\nu\rho}$ is the Levi--Civita symbol. These commutation relations show that the corresponding mode operators do not, in general, satisfy independent $\mathfrak{su}(2)$ algebras. Nevertheless, $[S_{z,\alpha},S_{z,\gamma}]=0$ for all $\alpha$ and $\gamma$. As a consequence, the quadratic contributions in Eq.~\eqref{twisting} are mutually commuting. The spectral decomposition therefore provides a set of commuting weighted collective interactions, whose spatial structure is determined by the eigenvectors of the underlying graph.

We now apply Eq.~\eqref{eq:spectral_katz} to several representative examples.

\begin{figure}[h]
    \centering   \includegraphics[width=\columnwidth]{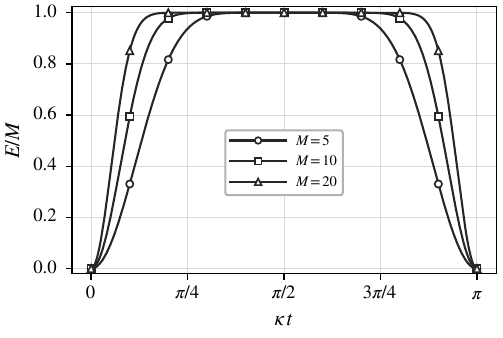}
    \caption{Entanglement Distance Eq.~\eqref{ED_compl} for $M=5,10,20$. At the time $\kappa t=\pi/2$, the state is LC-equivalent to $|\mathrm{GHZ}_5\rangle$, $|\mathrm{GHZ}_{10}\rangle$ and $|\mathrm{GHZ}_{20}\rangle$ respectively.}
    \label{ED_complete}
\end{figure}

\paragraph{Complete graph}
We first consider the complete graph $K_M$. In this case, all nodes are equivalent and all off-diagonal entries of the Katz kernel are equal. Indeed, for any vertex $i$, all
remaining vertices lie at distance one from $i$, and by symmetry they
all carry the same Katz coupling, which we denote by $\kappa$. The adjacency matrix of $K_M$ has one uniform eigenvector with components $u_{1,i}=1/\sqrt{M}$, $i=1,\dots,M$, with eigenvalue $\lambda_1=M-1$, while the remaining $M-1$ eigenvectors are orthogonal to $u_1$ and have eigenvalue $\lambda_\alpha=-1$ for $\alpha=2,\dots,M$. Therefore, using the completeness of the eigenvectors, one finds
\begin{equation}
\kappa
=
\frac{1}{\left[1-\beta(M-1)\right](1+\beta)} \, .
\end{equation}
In this case, the Hamiltonian Eq.~\eqref{twisting} reduces to the one-axis twisting Hamiltonian
\begin{equation}
    H=-J\kappa S_z^2 \, ,
\end{equation}
where $S_z=\sum_{i\in V}\sigma_z^{i}/2$. Therefore the complete graph $K_M$ generates a weighted graph state with a single entangling phase $\kappa t$ shared by all pairs of qubits. Thus the ED becomes
\begin{equation}
    E(t,\beta)
    =
    M\Big(1-
    \left[\cos(\kappa t)\right]^{2(M-1)}\Big)\, .
    \label{ED_compl}
\end{equation}
At the times $\kappa t^*=\pi/2+n\pi$, $n\in\mathbb{Z}$, the entanglement reaches its maximum value and the resulting state is
\begin{equation}
    |\psi(t^*)\rangle\sim_{\rm LC}
|\mathrm{GHZ}_M\rangle \, ,
\end{equation}
where $\sim_{LC}$ denotes equivalence under local Clifford operations \cite{hein2006entanglementgraphstatesapplications}. In the local case, $\beta=0$, although in the fermionic realization considered in Sec.~\ref{sec:model} the overall interaction scale $J$ vanishes, the Katz kernel reduces to the adjacency matrix, $K=A$, and the first maximum is reached at $t^*=\pi/2$. Thus, in this highly symmetric case, the full-time dynamics has a simple regular structure. The corresponding temporal profile is shown in Fig.(\ref{ED_complete}).

\paragraph{Star graph}
We now consider the star graph $S_M$, composed of one central vertex, which we label by $1$ for simplicity, and $M-1$ leaves. In this case the central
vertex and the leaves experience two distinct interaction patterns. The adjacency matrix satisfies $A^3=(M-1)A$. As a consequence, the Katz kernel can be written in closed form as
\begin{equation}
K=\frac{A+\beta A^2}{D} \, ,
\quad
D=1-(M-1)\beta^2 \, .
\end{equation}
The critical value for $\beta$ is $\beta_c=1/\sqrt{M-1}$. The relevant
off-diagonal entries of $K$ are $K_{1\ell}=1/D$ and
$K_{\ell m}=\beta/D$, with $\ell\neq m$ and $\ell,m=2,\dots,M$.
These couplings have a simple distance-shell interpretation. For the
central vertex, all $M-1$ leaves belong to the first shell and carry
the same coupling $1/D$. For a fixed leaf $\ell$, let
$\mathcal S_r(\ell)=\{j\in V:d(\ell,j)=r\}$ and
$n_\ell(r)=|\mathcal S_r(\ell)|$. The center is the unique vertex in
the first shell, $n_\ell(1)=1$, whereas the remaining $M-2$ leaves
form the second shell, $n_\ell(2)=M-2$. The corresponding Katz
couplings are $1/D$ and $\beta/D$, respectively. Since the star graph
has diameter two, no further distance shells occur, although longer walks still contribute to the Katz couplings within these shells.

The spectrum of the adjacency matrix contains two nonzero eigenvalues $\lambda_\pm=\pm \sqrt{M-1}$, with normalized eigenvectors $u_\pm=
\frac{1}{\sqrt{2}}
\left(
1,\pm \frac{1}{\sqrt{M-1}},
\dots,
\pm \frac{1}{\sqrt{M-1}}
\right)$, and remaining $M-2$ null eigenvalues with eigenvectors with zero component on the central vertex and leaf components whose sum vanishes. Accordingly, the Hamiltonian Eq.~\eqref{twisting} can be written equivalently as
\begin{equation}
H=-\frac{J}{D}\sigma_z^{1}S_z-\frac{J\beta}{D}S_z^2,
\end{equation}
where $S_z=\sum_{i\in V\backslash\{1\}}\sigma_z^i/2$. 

Equation~\eqref{eq:exact_Ei} then gives for the central vertex
\begin{equation}
E_1(t,\beta)
=
1-
\bigg[
\cos^2\Big(\frac{t}{D}\Big)
\bigg]^{M-1}\, ,
\end{equation}
whereas for any leaf $\ell=2,\dots,M$ one finds
\begin{equation}
E_\ell(t,\beta)
=
1-
\cos^2\Big(\frac{t}{D}\Big)
\left[
\cos^2\Big(\frac{\beta t}{D}\Big)
\right]^{M-2}\, .
\end{equation}
The powers appearing in these expressions have a direct distance-shell
origin. For the central vertex, the exponent $M-1$ counts the leaves in
its only distance shell. For a leaf, the first factor comes from the
single center at distance one, while the exponent $M-2$ counts the
vertices in the second shell, all of which carry the same
walk-mediated coupling $\beta/D$. The central-vertex contribution therefore contains a single Katz scale,
whereas the leaf contribution contains both the direct center--leaf
scale $1/D$ and the indirect leaf--leaf scale $\beta/D$.

\begin{figure}[h]
    \centering   \includegraphics[width=\columnwidth]{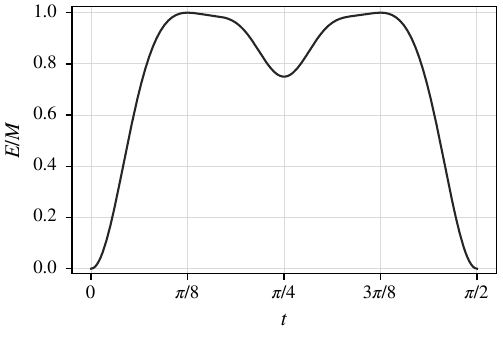}
    \caption{Entanglement Distance for the star graph with $M=4$ and $q=2$, corresponding to $\beta=1/2$. At the time $t_2^*=\pi/4$, the state is LC-equivalent to $|+\rangle_1|\mathrm{GHZ}_3\rangle$.}
\label{ED_star_even}
\end{figure}

\begin{figure}[h]
    \centering   \includegraphics[width=\columnwidth]{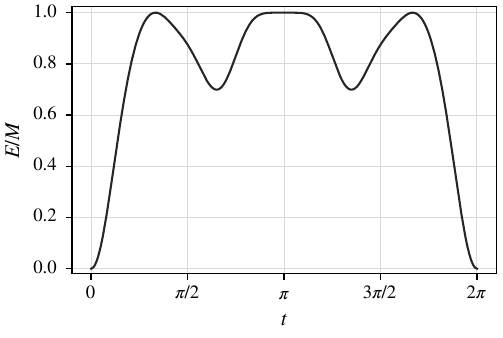}
    \caption{Entanglement Distance for the star graph with $M=4$ and $q=3$ corresponding to $\beta=1/3$. At the time $t_2^*=\pi$, the state is LC-equivalent to $|\mathrm{GHZ}_4\rangle$.}
\label{ED_star_odd}
\end{figure}

At the first special time
$t_1^*=\pi D/2$, the entanglement reaches its maximum value and the evolved state is
\begin{equation}
|\psi(t_1^*)\rangle
=
\exp\left[
i\frac{\beta\pi}{2}S_z^2
\right]
|\mathrm{Star}_M\rangle 
\end{equation}
where $|\mathrm{Star}_M\rangle$ is the standard graph state associated with the star graph. Since
$
|\mathrm{Star}_M\rangle
\sim_{\rm LC}
|\mathrm{GHZ}_M\rangle
$,
the state at this first special time produces a GHZ-like state twisted by a collective interaction acting only on the leaves.
At the second special time $t_2^*=\pi D/(2\beta)$, the leaves subsystem is maximally entangled. The full resulting state is
\begin{equation}
|\psi(t_2^*)\rangle=
\exp\left[
i\frac{\pi}{2\beta}
\sigma_z^{1} S_z
\right]
|+\rangle_1
|K_{M-1}\rangle_{\rm leaves}\, .
\end{equation}
Since $|K_{M-1}\rangle_{\rm leaves}
\sim_{\rm LC}
|\mathrm{GHZ}_{M-1}\rangle$, this state can be interpreted as a GHZ state on the leaves, dressed by a collective Z-twist coupling the leaves to the center.
If $1/\beta=q\in\mathbb Z$, with $q>\sqrt{M-1}$, the remaining center-leaf phase is also Clifford. For even $q$, the center-leaf gates are locally trivial and
\begin{equation}
|\psi(t_2^*)\rangle
\sim_{\rm LC}
|+\rangle_1
|K_{M-1}\rangle_{\rm leaves}
\sim_{\rm LC}
|+\rangle_1|\mathrm{GHZ}_{M-1}\rangle \, .
\end{equation}
For odd $q$, the center is connected to all leaves by Clifford entangling gates and the resulting graph is the complete graph $K_M$. Hence
\begin{equation}
|\psi(t_2^*)\rangle
\sim_{\rm LC}
|K_M\rangle
\sim_{\rm LC}
|\mathrm{GHZ}_M\rangle \, .
\end{equation}
In Fig. (\ref{ED_star_even}) and (\ref{ED_star_odd}) we report the entanglement profile for $M=4$ and $q=2$ and $q=3$ respectively.

\paragraph{Path Graph}
A different situation is obtained for the path graph $P_M$, a 1D chain with open boundary conditions, whose
adjacency matrix is
$A_{ij}=\delta_{i,j+1}+\delta_{i+1,j}$,
with $i,j=1,\dots,M$. Unlike the complete and star graphs, the
interaction pattern seen by a vertex now depends on its position along
the chain. The path also provides a direct realization of the distance
hierarchy discussed above, since two vertices at distance $r$ are
connected by a unique shortest path. Its adjacency matrix is analytically diagonalizable, with eigenvalues
and normalized eigenvectors
\begin{equation}
    \lambda_\alpha
    =
    2\cos\bigg(\frac{\pi\alpha}{M+1}\bigg) \, ,
    \,
    u_{\alpha,j}
    =
    \sqrt{\frac{2}{M+1}}
    \sin\bigg(\frac{j\pi\alpha}{M+1}\bigg) \, ,
\end{equation}
with $\alpha=1,\dots,M$. In contrast with the previous examples, the Katz couplings are not
fixed solely by the distance between two vertices: finite boundaries
also retain information on their absolute positions along the chain.
As a consequence, boundary and bulk vertices generally experience
different interaction patterns.

For finite $M$, a useful explicit form follows from the inverse of the tridiagonal matrix $I-\beta A(P_M)$. Writing $D_M=\det[I-\beta A(P_M)]$, with $D_0=D_1=1$ and $D_M=D_{M-1}-\beta^2D_{M-2}$, one obtains, for $i\neq j$,
\begin{equation}
\label{eq:path_katz_closed}
K_{ij}
=
\beta^{b-a-1}
\frac{
D_{a-1}D_{M-b}
}{
D_M
} \, ,
\end{equation}
where $a=\min(i,j)$ and $b=\max(i,j)$. For a fixed vertex $i$, the $r$th distance shell is
$\mathcal S_r(i)=\{j\in V:|i-j|=r\}$, with population
$n_i(r)=|\mathcal S_r(i)|$. The shell contains the two vertices
$i-r$ and $i+r$ whenever both belong to the chain, so that
$n_i(r)=2$, whereas $n_i(r)=1$ when only one of them is present.
Since the path contains a unique shortest path between any pair of
vertices, Eq.~\eqref{eq:katz_distance_order} implies that a vertex in
$\mathcal S_r(i)$ first couples to $i$ with strength
$\beta^{r-1}$. Because the path is bipartite, the next allowed walks have length $r+2$, so that 
\begin{equation}
    K_{ij}
    =
    \beta^{|i-j|-1}
    +
    O\!\left(\beta^{|i-j|+1}\right) \, .
    \label{eq:path_distance_hierarchy}
\end{equation}
Successive distance shells thus enter the Katz interaction at successive
powers of $\beta$. Unlike the star graph, however, a distance shell does not in general
correspond to a single Katz coupling. For a finite path, the two
vertices $i-r$ and $i+r$ can have different couplings to $i$ because
the number of longer walks connecting them to $i$ depends on their
position relative to the boundaries.

To make the spatial decay of the Katz couplings explicit, we introduce
$
q=e^{-1/\xi}
=
2\beta/(1+\sqrt{1-4\beta^2})$, $
\Delta_\beta=\sqrt{1-4\beta^2}$,
for $\beta<1/2$. Using the closed form of the determinants $D_M$,
the exact finite-size kernel in Eq.~\eqref{eq:path_katz_closed} can be
rewritten, for $i<j$, as
\begin{equation}
K_{ij}
=
C e^{-(j-i)/\xi}
\left(1-e^{-2i/\xi}\right)
\left(1-e^{-2(M-j+1)/\xi}\right),
\label{eq:path_exponential_kernel}
\end{equation}
where
\begin{equation}
C=
\left[
\beta\Delta_\beta\left(1-q^{2(M+1)}\right)
\right]^{-1}   \, .
\end{equation}
The case $i>j$ follows from the symmetry $K_{ij}=K_{ji}$.
For a fixed vertex $i$, this gives the two profiles
\begin{equation}
\begin{aligned}
K_{i,i+d}
&=
C e^{-d/\xi}
\left(1-e^{-2i/\xi}\right)
\left(1-e^{-2(M-i-d+1)/\xi}\right)\, ,
\\[1.20em]
K_{i,i-d}
&=
C e^{-d/\xi}
\left(1-e^{-2(i-d)/\xi}\right)
\left(1-e^{-2(M-i+1)/\xi}\right)\, ,
\label{eq:path_right_left_profiles}
\end{aligned}
\end{equation}
with $d=1,\dots,M-i$ in the first line and $d=1,\dots,i-1$
in the second one. For a fixed vertex $i$, these two expressions
resolve the $d$th distance shell into its right and left couplings.
In the bulk and far from the boundaries they become equivalent,
whereas finite-size effects distinguish them as either endpoint of the
chain is approached.
In the regime $M\gg\xi$, and for vertices sufficiently far from both
boundaries, the finite-size factors approach unity and the two profiles
reduce to the same exponential form,
\begin{equation}
K_{i,i\pm d}
\simeq
\frac{1}{\beta\Delta_\beta}e^{-d/\xi} \, .
\end{equation}

The ratio $\xi/M$ controls the relevant regime. If $\xi\ll M$,
the factor $e^{-d/\xi}$ suppresses distant vertices exponentially, and
the exact ED product is effectively controlled by a local neighborhood
of $i$. Indeed, a vertex at distance $d$ from $i$ can first contribute
appreciably when the phase $tK_{ij}$ becomes of order unity. Writing
$K_{ij}\sim K_i^{\rm loc}e^{-d/\xi}$, where $K_i^{\rm loc}$ denotes
the local prefactor of the exponential profile, this defines the
characteristic scale
$t_d\sim e^{d/\xi}/K_i^{\rm loc}$.
Thus increasingly distant distance shells become dynamically relevant
only at exponentially larger values of the entangling parameter. Equivalently, at fixed $t$ we define the effective distance $d_t$ as
the scale at which $tK_{ij}$ becomes of order unity. From
$tK_i^{\rm loc}e^{-d_t/\xi}\sim1$, one obtains
$d_t\sim\xi\log(tK_i^{\rm loc})$, for
$tK_i^{\rm loc}\gtrsim1$. The local ED contribution can therefore be
approximated as
\begin{equation}
E_i(t,\beta)
\simeq
1-
\prod_{|i-j|\lesssim d_t}
\cos^2\left(tK_{ij}\right) \, ,
\quad
d_t\sim\xi\log\left(tK_i^{\rm loc}\right)\, .
\end{equation}
Therefore, in this regime the ED is controlled by a finite neighborhood
of the vertex, whose effective size grows only logarithmically with
$t$.

The crossover occurs when $\xi\sim M$. Taking $\xi=M$ as a reference
scale gives
$\beta_*(M)=1/[2\cosh(1/M)]
\simeq1/2-1/(4M^2)$.
At this scale, distance shells extending over a finite fraction of the
chain are no longer exponentially suppressed. Equivalently, vertices
at distances of order $M$ no longer require exponentially large values
of $t$ to contribute appreciably. In this regime, no local truncation
of the ED product is natural, and the full finite-size expression for
the Katz kernel must be retained.

\begin{figure}[h]
    \centering   \includegraphics[width=\columnwidth]{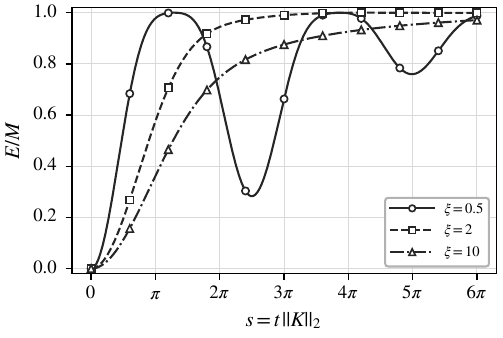}
    \caption{Entanglement Distance of the path graph with $M=20$ and $\xi=0.5, 2.0, 10.0$, as a function of the rescaled time $s=t\lVert K\rVert_2$.}
\label{path1}
\end{figure}

Finally, when $\xi\gg M$, the exponential factor no longer produces a
hierarchy of distances inside the chain, since
$e^{-|i-j|/\xi}\simeq1$ for all pairs of vertices. This regime
corresponds to
\begin{equation}
    \frac{1}{2}-\beta\ll\frac{1}{4M^2} \, .
\end{equation}
Expanding the remaining factors in Eq.~\eqref{eq:path_exponential_kernel}, one obtains
\begin{equation}
K_{ij}
\simeq
\frac{
4\min(i,j)[M-\max(i,j)+1]
}{
M+1
}\, .
\end{equation}
Thus the Katz interaction is no longer effectively localized around
$i$, but extends over the whole finite chain. Couplings to distant
vertices are no longer exponentially suppressed, and the local ED
contribution takes the form
\begin{equation}
E_i(t,\beta)
\simeq
1-
\prod_{j\neq i}
\cos^2\left[
t
\frac{
4\min(i,j)[M-\max(i,j)+1]
}{
M+1
}
\right] \, .
\end{equation}
This is the opposite limit to $\xi\ll M$: instead of a product
effectively restricted to a finite neighborhood, the ED involves
couplings extending across the whole chain. To compare the dynamics at different values of $\xi$ on the same interaction timescale, we introduce the rescaled time $s=t\lVert K\rVert_2$ where $\lVert K\rVert_2$ denotes the spectral norm of the Katz kernel. Since $K=A(I-\beta A)^{-1}$ is real and symmetric, its spectral norm is given by $\lVert K\rVert_2=\max_{\alpha}\left|
k_{\alpha}
\right|$. Figure (\ref{path1}) shows the Entanglement Distance as a function of $s$ for representative values of $\xi$.

\paragraph{Frucht graph}
As a final example, we consider the Frucht graph $F$ shown in Fig. (\ref{fig:frucht_graph}). The Frucht graph is a cubic asymmetric graph on $M=12$ vertices \cite{Frucht_1949}. This example is particularly
useful because all vertices have the same degree, while their extended
graph environments are not equivalent. It is finite and sparse, but unlike the previous examples it does not possess nontrivial vertex symmetries that reduce the exact-time dynamics to a small set of
equivalent Katz couplings.
For this graph no compact analytic reduction of the kernel is available, in contrast to the previous cases. However, the largest eigenvalue is $\lambda_1=3$. For this graph no compact analytic reduction of the kernel is available,
in contrast to the previous cases. However, since the graph is
$3$-regular, its largest adjacency eigenvalue is $\lambda_1=3$. We
therefore compute the Katz kernel numerically and evaluate the total ED
from Eq.~\eqref{exact}. The Frucht graph thus provides a deterministic
example in which the Katz-mediated entanglement dynamics is governed by
a genuinely vertex-dependent pattern of couplings, and therefore by a
much less reducible set of exact-time entangling scales.
Figure~\ref{ED_frucht_1} shows the entanglement profile for two
representative values of $\beta$ as a function of the rescaled time
$s=t\lVert K\rVert_2$.
\begin{figure}[t]
\centering
\begin{tikzpicture}[
    scale=0.6,
    fruchtedge/.style={black, line width=0.55pt, line cap=round},
    fruchtnode/.style={
        circle,
        draw=black,
        line width=0.15pt,
        fill=black,
        minimum size=1.2mm,
        inner sep=1.2pt
    }
]

\coordinate (a) at (-1.17,  1.58);
\coordinate (b) at ( 1.77,  1.31);
\coordinate (c) at ( 1.47,  0.47);
\coordinate (d) at ( 2.23,  0.27);
\coordinate (e) at (-0.62,  0.21);
\coordinate (f) at ( 0.63, -0.17);
\coordinate (g) at (-2.23, -0.19);
\coordinate (h) at (-1.47, -0.41);
\coordinate (i) at ( 0.98, -1.17);
\coordinate (j) at (-1.80, -1.24);
\coordinate (k) at ( 1.75, -1.43);
\coordinate (l) at ( 0.77, -2.05);

\draw[fruchtedge] (a) to[out=25,in=155] (b);
\draw[fruchtedge] (a) -- (e);
\draw[fruchtedge] (a) to[out=200,in=105] (g);

\draw[fruchtedge] (b) -- (c);
\draw[fruchtedge] (b) to[out=-35,in=70] (d);

\draw[fruchtedge] (c) -- (d);
\draw[fruchtedge] (c) -- (f);

\draw[fruchtedge] (d) to[out=-70,in=35] (k);

\draw[fruchtedge] (e) -- (f);
\draw[fruchtedge] (e) -- (h);

\draw[fruchtedge] (f) -- (i);

\draw[fruchtedge] (g) -- (h);
\draw[fruchtedge] (g) to[out=-95,in=175] (j);

\draw[fruchtedge] (h) -- (j);

\draw[fruchtedge] (i) -- (k);
\draw[fruchtedge] (i) -- (l);

\draw[fruchtedge] (j) to[out=-25,in=-160] (l);

\draw[fruchtedge] (k) to[out=-145,in=-15] (l);

\node[fruchtnode] at (a) {};
\node[fruchtnode] at (b) {};
\node[fruchtnode] at (c) {};
\node[fruchtnode] at (d) {};
\node[fruchtnode] at (e) {};
\node[fruchtnode] at (f) {};
\node[fruchtnode] at (g) {};
\node[fruchtnode] at (h) {};
\node[fruchtnode] at (i) {};
\node[fruchtnode] at (j) {};
\node[fruchtnode] at (k) {};
\node[fruchtnode] at (l) {};

\end{tikzpicture}
\caption{Frucht graph $F$, used as a deterministic asymmetric graph on $12$ vertices.}
\label{fig:frucht_graph}
\end{figure}
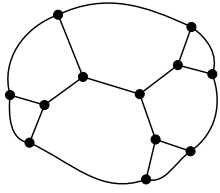
\begin{figure}[h]
    \centering   \includegraphics[width=\columnwidth]{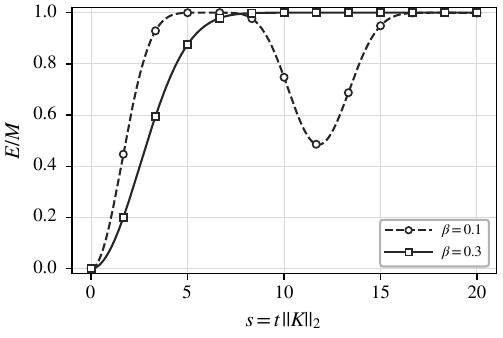}
    \caption{Entanglement Distance for the Frucht graph with $\beta=0.1$ and $\beta=0.3$}
    \label{ED_frucht_1}
\end{figure}

The examples above show that, although the ED is known exactly, its
finite-time dynamics depends strongly on how the structure of the Katz
kernel is organized by the underlying graph. In the complete graph,
symmetry reduces the dynamics to a single Katz coupling. In the star
graph, the central vertex and the leaves probe two distinct
distance-shell structures and hence two different coupling patterns.
For the path graph, successive distance shells enter through a
hierarchy of couplings whose exact values also retain the effect of the
boundaries. Finally, in the Frucht graph the absence of nontrivial
vertex symmetries prevents such a reduction, and the dynamics involves
a genuinely vertex-dependent pattern of Katz couplings. Since the ED
contains products of factors $\cos^2(tK_{ij})$, these different
couplings enter the dynamics nonlinearly and generate graph-dependent
oscillatory structures.

We now move from these graph-specific examples to the general behavior
of the entanglement as the Katz interaction extends through progressively
longer walks. In the weak-propagation regime, the dynamics can be
organized perturbatively in $\beta$ without expanding in the entangling
parameter $t$, thereby retaining the full time dependence within the
controlled perturbative window. This allows the first structural
corrections to the original graph interaction to be identified directly.
At the opposite end, as $\beta$ approaches the convergence threshold
$\beta_c=1/\lambda_1$, the Katz kernel becomes increasingly dominated
by the principal spectral mode. These two limits provide complementary
descriptions of the entanglement dynamics across short- and long-range
propagation.

\section{Entanglement across propagation regimes}
\label{sec:propagation}

The exact expression in Eq.~\eqref{eq:exact_Ei} shows that two
ingredients determine the entanglement dynamics. The entangling
parameter $t$ controls the accumulated interaction phase, whereas the
Katz kernel $K$ determines how the interaction is distributed over the
graph. These two effects need not be varied simultaneously. We therefore retain the full
$t$ dependence and organize the Katz interaction perturbatively in
$\beta$, whose convergence threshold is $\beta_c=1/\lambda_1$.

The two limiting regimes have qualitatively different structures. For
$\beta\ll\beta_c$, the first corrections to the adjacency interaction
arise from the shortest indirect walks and can be related directly to
local graph motifs. In the opposite limit $\beta\to\beta_c^{-}$, longer
walks are progressively enhanced and the Katz kernel becomes
increasingly dominated by the principal adjacency mode. We analyze
these two regimes separately below.

\subsection{Weak-propagation regime for fixed graphs}

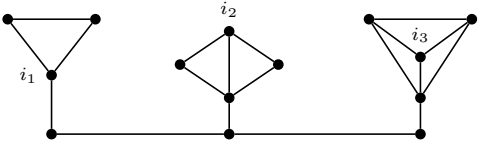
\begin{figure}[t]
    \centering
    \begin{tikzpicture}[line width=0.6pt]


        \node[
            draw, fill=black, circle, inner sep=1.2pt,
            label={[font=\scriptsize]left:$i_1$}
        ] (i1) at (0.00,0.78) {};

        \node[draw, fill=black, circle, inner sep=1.2pt]
            (a1) at (-0.58,1.52) {};

        \node[draw, fill=black, circle, inner sep=1.2pt]
            (a2) at ( 0.58,1.52) {};

        \node[draw, fill=black, circle, inner sep=1.2pt]
            (x1) at (0.00,0.00) {};

        \draw (i1)--(a1);
        \draw (i1)--(a2);
        \draw (a1)--(a2);
        \draw (i1)--(x1);


        \node[
            draw, fill=black, circle, inner sep=1.2pt,
            label={[font=\scriptsize]above:$i_2$}
        ] (i2) at (2.35,1.36) {};

        \node[draw, fill=black, circle, inner sep=1.2pt]
            (b1) at (1.70,0.92) {};

        \node[draw, fill=black, circle, inner sep=1.2pt]
            (b2) at (3.00,0.92) {};

        \node[draw, fill=black, circle, inner sep=1.2pt]
            (b3) at (2.35,0.48) {};

        \node[draw, fill=black, circle, inner sep=1.2pt]
            (x2) at (2.35,0.00) {};

        \draw (i2)--(b1);
        \draw (i2)--(b2);
        \draw (i2)--(b3);

        \draw (b1)--(b3);
        \draw (b2)--(b3);

        \draw (b3)--(x2);


        \node[draw, fill=black, circle, inner sep=1.2pt]
            (c1) at (4.20,1.52) {};

        \node[draw, fill=black, circle, inner sep=1.2pt]
            (c2) at (5.56,1.52) {};

        \node[draw, fill=black, circle, inner sep=1.2pt]
            (c3) at (4.88,0.48) {};

        \node[
            draw, fill=black, circle, inner sep=1.2pt,
            label={[font=\scriptsize,label distance=0.1pt]above:$i_3$}
        ] (i3) at (4.88,1.02) {};

        \node[draw, fill=black, circle, inner sep=1.2pt]
            (x3) at (4.88,0.00) {};

        \draw (c1)--(c2);
        \draw (c1)--(c3);
        \draw (c2)--(c3);

        \draw (i3)--(c1);
        \draw (i3)--(c2);
        \draw (i3)--(c3);

        \draw (c3)--(x3);


        \draw (x1)--(x2);
        \draw (x2)--(x3);

    \end{tikzpicture}

    \caption{
    Connected graph with three labeled vertices of equal degree,
    $d_{i_1}=d_{i_2}=d_{i_3}=3$, and distinct local structures.
    }
    \label{graph_single_site_weak}
\end{figure}

\begin{figure}[h]
    \centering   
    \includegraphics[width=\columnwidth]{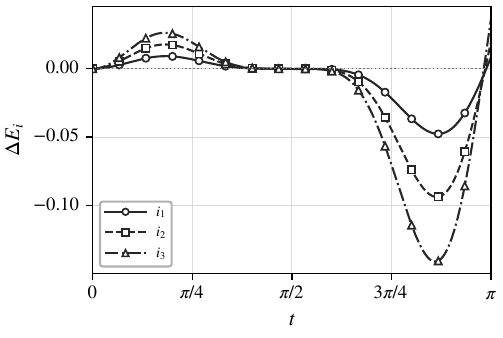}
    \caption{Weak-propagation correction
$\Delta E_i(t,\beta)=E_i(t,\beta)-E_i(t,0)$
for the three selected vertices of the graph in
Fig.~\ref{graph_single_site_weak}, with
$\beta=\eta\beta_c$ and $\eta=0.05$.
All three vertices have degree $d_i=3$, and therefore share the same
zeroth-order contribution $E_i(t,0)=1-\cos^6 t$.
Their local structures differ as
$(T_i,W_i,Q_i)=(1,12,2)$ for $i_1$,
$(2,16,6)$ for $i_2$, and
$(3,22,12)$ for $i_3$.
The curves are obtained from the exact expression in Eq.~\eqref{eq:exact_Ei}. The time range $0\leq t\leq\pi$ is chosen consistently with the
weak-propagation condition $t\beta\lambda_1^2\ll1$ for the parameters
considered. }
    \label{ED_single_site_weak}
\end{figure}

We first consider the regime $\beta\ll\beta_c$. The Katz kernel is then
\begin{equation}
    K=A+\beta A^2+\beta^2A^3+O(\beta^3),
    \label{eq:weak_katz_kernel}
\end{equation}
so that increasing orders in $\beta$ progressively incorporate longer
walks. Since the entanglement phases are proportional to $tK_{ij}$,
the weak-propagation expansion also requires that the correction to the
adjacency interaction does not accumulate a large phase. A sufficient
condition is $t\,\|K-A\|_2=t\,\frac{\beta\lambda_1^2}{1-\beta\lambda_1}\ll 1$, which, for $\beta\lambda_1\ll 1$ reduces to $t\beta\lambda_1^2\ll 1$. Within this time window, expanding Eq.~\eqref{eq:exact_Ei} only in
$\beta$ gives
\begin{equation}
\begin{aligned}
E_i(t,\beta)
={}&
1-\cos^{2d_i}t
+
4\beta tT_i\tan t\,\cos^{2d_i}t
\\
&+
\beta^2\cos^{2d_i}t
\bigg[
2t\tan t\,W_i
+t^2\left(W_i-d_i^2\right)
\\
&
+t^2\tan^2t
\left(Q_i-8T_i^2\right)
\bigg]
+O(\beta^3),
\end{aligned}
\label{eq:weak_propagation_Ei}
\end{equation}
where
$T_i=\frac{1}{2}(A^3)_{ii}
=\frac{1}{2}\sum_{j,k\in V}A_{ij}A_{jk}A_{ki}$
is the number of triangles containing vertex $i$,
$W_i=(A^4)_{ii}$, and
$Q_i=\sum_{j\in\mathcal N(i)}[(A^2)_{ij}]^2$.
The first-order term has a direct interpretation. For an existing edge
$(i,j)$, $(A^2)_{ij}$ is the number of common neighbors of $i$ and
$j$, namely the number of triangles sharing that edge. Summing over the
neighbors of $i$ gives
$\sum_{j\in\mathcal N(i)}(A^2)_{ij}=2T_i$, which explains the
appearance of the local triangle count $T_i$ at order $\beta$.
Couplings between vertices that are not directly connected start instead
at order $\beta$, and their contribution to the ED is therefore
quadratic in $\beta$. At second order, four-step closed structures become relevant. In
particular,
$
W_i=(A^4)_{ii}
=
d_i^2+\sum_{j\in\mathcal N(i)}(d_j-1)+2C_{4,i},
$
where $C_{4,i}$ is the number of simple four-cycles containing $i$.
Thus $W_i$ contains both the local degree structure around $i$ and the
four-cycle contribution. The quantity
$Q_i=\sum_{j\in\mathcal N(i)}[(A^2)_{ij}]^2$ instead measures the
multiplicity of triangles sharing the edges incident on $i$; explicitly,
$
Q_i
=
2T_i+
2\sum_{j\in\mathcal N(i)}
\binom{(A^2)_{ij}}{2}.
$
The term proportional to $T_i^2$ results from the nonlinear product
structure of the ED. We illustrate this hierarchy using the graph shown in
Fig.~\ref{graph_single_site_weak}. To isolate the contribution generated
by weak propagation, in Fig.~\ref{ED_single_site_weak} we plot $\Delta E_i(t,\beta)
=
E_i(t,\beta)-E_i(t,0)$, where $E_i(t,0)=1-\cos^6 t$ is the same for the three selected vertices.

Summing over all vertices gives
\begin{equation}
\begin{aligned}
E(t,\beta)
={}&
M-\sum_{i\in V}\cos^{2d_i}t
+
4\beta t\tan t
\sum_{i\in V}T_i\cos^{2d_i}t
\\
&+
\beta^2
\sum_{i\in V}\cos^{2d_i}t
\bigg[
2t\tan t\,W_i
+t^2\left(W_i-d_i^2\right)
\\
&
+t^2\tan^2t
\left(Q_i-8T_i^2\right)
\bigg]
+O(\beta^3).
\end{aligned}
\label{eq:weak_propagation_global}
\end{equation}
At finite $t$, each local term is weighted by
$\cos^{2d_i}t$, so triangles, four-cycles, the degree structure of the
neighborhood, and overlapping triangles contribute differently
depending on the vertices on which they occur. The finite-time
dynamics therefore retains information on the distribution of these
structures across the graph, rather than only on their total number.

Equation~\eqref{eq:weak_propagation_global} therefore provides a
finite-time motif expansion of the ED in the weak-propagation regime.
Triangles enter at first order in $\beta$, while the second-order term
resolves four-cycles, the local degree structure, and overlapping
triangles.

\subsection{Strong-propagation regime}

We now consider the opposite limit corresponding to $\beta\to\beta_c^{-}$. From the spectral representation of the Katz
kernel, the principal coefficient
$k_1=\lambda_1/(1-\beta\lambda_1)$ diverges at the convergence
threshold, while the remaining coefficients stay finite. The kernel is
therefore asymptotically dominated by the principal adjacency mode,
$K\simeq k_1u_1u_1^T$.

At fixed $t$, the phases $tKij$ become arbitrarily large in
this limit and the ED has no smooth limiting profile. We therefore
keep fixed the accumulated principal-mode phase $s=tk_1$. In the limit
$\beta\to\beta_c^{-}$ at fixed $s$, Eq.~\eqref{eq:exact_Ei} becomes
\begin{equation}
E_i(s)
\simeq
1-
\prod_{j\neq i}
\cos^2\left(su_{1,i}u_{1,j}\right).
\label{eq:strong_propagation_Ei}
\end{equation}
The spatial structure of $u_1$ therefore determines how the
strong-propagation entanglement is distributed over the graph. If the
principal eigenvector is delocalized over $M$ vertices,
$u_{1,i}^2\sim1/M$ and
$\sum_i u_{1,i}^4\sim1/M$. The phases in
Eq.~\eqref{eq:strong_propagation_Ei} are then distributed over the
whole graph, with each pairwise phase scaling as $s/M$ for a fully
delocalized mode.
If instead $u_1$ is localized over $m\ll M$ vertices, then
$\sum_i u_{1,i}^4\sim1/m$ and its components are appreciable only
within the localization region. The phases involving vertices outside
this region are correspondingly suppressed, and the entanglement
dynamics is concentrated on the vertices supporting the principal mode.

\subsection{Random graphs in the weak-propagation regime}

We now extend the weak-propagation result to Erd\H{o}s--R\'enyi random graph ensemble. For statistically equivalent vertices, we propose to consider the ensemble-averaged
ED density
\begin{equation}
    e(t,\beta)
    =
    \frac{1}{M}\,
    \mathbb{E}\!\left[E(t,\beta)\right].
\end{equation}
Let $P(d)$ denote the degree distribution and introduce the conditional
averages
$\tau_d=\mathbb{E}[T_i|d_i=d]$,
$\omega_d=\mathbb{E}[W_i|d_i=d]$,
$\eta_d=\mathbb{E}[Q_i|d_i=d]$, and
$\theta_d=\mathbb{E}[T_i^2|d_i=d]$.
Setting $x=\cos^2t$, Eq.~\eqref{eq:weak_propagation_global} gives
\begin{align}
e(t,\beta)
={}&
1-G_0(x)
+
4\beta t\tan t\,G_T(x)
\nonumber\\
&+
\beta^2
\bigg[
2t\tan t\,G_W(x)
+
t^2\left[
G_W(x)-D_2(x)
\right]
\nonumber\\
&
+
t^2\tan^2t
\left[
G_Q(x)-8G_{T^2}(x)
\right]
\bigg]
+
O(\beta^3),
\label{eq:random_weak_general}
\end{align}
where
$G_0(x)=\sum_dP(d)x^d$, $G_T(x)=\sum_dP(d)\tau_d x^d$,
$G_W(x)=\sum_dP(d)\omega_d x^d$, $G_Q(x)=\sum_dP(d)\eta_d x^d$, $G_{T^2}(x)=\sum_dP(d)\theta_d x^d$, and $D_2(x)=\sum_dP(d)d^2x^d=xG_0'(x)+x^2G_0''(x)$.
The zeroth-order dynamics is completely determined by the degree
distribution through $G_0(x)$, while the first Katz correction depends
on the degree-resolved triangle statistics through $G_T(x)$. At second
order, $G_W(x)$ introduces the neighborhood-degree and four-cycle
contributions contained in $(A^4)_{ii}$, whereas $G_Q(x)$ and
$G_{T^2}(x)$ retain the information associated with triangles sharing
the same edges and with their nonlinear combination in the ED.
At finite $t$, all these quantities are weighted by $x^d=\cos^{2d}t$.
The ensemble-averaged dynamics is therefore sensitive not only to the
mean abundance of the corresponding graph motifs, but also to how they
are distributed among vertices of different degree. At the same time, the weak-propagation condition must hold realization by
realization. Since $\beta_c=1/\lambda_1$ fluctuates across the ensemble,
we assume $\beta\lambda_1\ll1$ for the relevant realizations.

\subsubsection{Erd\H{o}s--R\'enyi graphs}
We first consider the Erd\H{o}s--R\'enyi ensemble $G(M,p)$, for which
\begin{equation}
    P(d)
    =
    \binom{M-1}{d}p^d(1-p)^{M-1-d}.
\end{equation}
Conditioned on $d_i=d$, the edges among the $d$ neighbors of $i$
remain independent with probability $p$. The number of triangles
incident on $i$ is therefore binomial, which gives
\begin{equation}
    \tau_d
    =
    p\binom{d}{2}
    \label{eq:ER_tau}.
\end{equation}
The remaining conditional averages entering
Eq.~\eqref{eq:random_weak_general} can also be evaluated explicitly.
Using
\[
W_i
=
d_i^2
+
\sum_{j\in\mathcal N(i)}(d_j-1)
+
2C_{4,i},
\]
one finds
\begin{equation}
    \omega_d
    =
    d^2
    +
    d(M-2)p
    +
    d(d-1)(M-3)p^2 .
    \label{eq:ER_omega}
\end{equation}
The last term is the contribution of four-cycles: for each pair of
neighbors of $i$, any of the remaining $M-3$ vertices can close a
four-cycle with probability $p^2$.
For $Q_i$, the number $(A^2)_{ij}$ of triangles sharing a given edge
$(i,j)$ is binomial with parameters $d-1$ and $p$. Hence
\begin{equation}
    \eta_d
    =
    d\left[
    (d-1)p(1-p)
    +(d-1)^2p^2
    \right]
    =
    p(d)_2+p^2(d)_3 ,
    \label{eq:ER_eta}
\end{equation}
where $(d)_r=d(d-1)\cdots(d-r+1)$ denotes the falling factorial.
Similarly, since
$T_i\sim{\rm Bin}\!\left(\binom{d}{2},p\right)$,
\begin{equation}
    \theta_d
    =
    p(1-p)\binom{d}{2}
    +
    p^2\binom{d}{2}^{\!2}.
    \label{eq:ER_theta}
\end{equation}
Writing $n=M-1$ and $a(x)=1-p+px$, the binomial identity
\begin{equation}
    \sum_d P(d)(d)_r x^d
    =
    (n)_r p^r x^r a(x)^{n-r}
\end{equation}
gives all the generating functions entering
Eq.~\eqref{eq:random_weak_general}.
\begin{align}
G_0(x)
&=
a^n ,
\label{eq:G_0}
\\
G_T(x)
&=
\frac{(n)_2}{2}\,
p^3x^2a^{n-2},
\label{eq:G_T}
\\
G_W(x)
&=
np\,x\,a^{n-1}\left[1+(n-1)p\right]
\nonumber\\
&\quad
+(n)_2p^2x^2a^{n-2}
\left[1+(n-2)p^2\right],
\label{eq:G_W}
\\
D_2(x)
&=
np\,x\,a^{n-1}
+
(n)_2p^2x^2a^{n-2},
\label{eq:D_2}
\\
G_Q(x)
&=
(n)_2p^3x^2a^{n-2}
+
(n)_3p^5x^3a^{n-3},
\label{eq:G_Q}
\\
G_{T^2}(x)
&=
\frac{(n)_2}{2}p^3x^2a^{n-2}
+
(n)_3p^5x^3a^{n-3}
\nonumber\\
&\quad
+
\frac{(n)_4}{4}p^6x^4a^{n-4}.
\label{eq:G_T2}
\end{align}
Equations~\eqref{eq:ER_tau}--\eqref{eq:G_T2}
therefore provide the finite-$M$ weak-propagation ED for the
Erd\H{o}s--R\'enyi ensemble through $O(\beta^2)$. In this case the
degree, triangle, four-cycle, and overlapping-triangle contributions
are all fixed by the two ensemble parameters $M$ and $p$.

\paragraph{Sparse Erd\H{o}s--R\'enyi regime.}
We first consider the sparse regime $p=c/M$
with $c=O(1)$, so that the mean degree approaches $c$ as
$M\to\infty$. In this regime, the largest eigenvalue $\lambda_1$ depends on the specific graph realization and cannot in general be expressed solely in terms of $c$. For a fixed finite $t$, we therefore
choose $\beta$ realization by realization such that $\beta\lambda_1\ll1$ and $t\beta\lambda_1^2\ll 1$. The degree distribution becomes Poissonian and
\begin{equation}
    G_0(x)
    \longrightarrow
    e^{c(x-1)}.
\end{equation}
In the same limit,
\begin{equation}
    G_T(x),\,G_Q(x),\,G_{T^2}(x)
    =
    O(M^{-1}),
\end{equation}
whereas
\begin{align}
    G_W(x)
    &\longrightarrow
    e^{c(x-1)}
    \left[
        cx+c^2x(1+x)
    \right],
    \\
    D_2(x)
    &\longrightarrow
    e^{c(x-1)}
    \left[
        cx+c^2x^2
    \right].
\end{align}
Substituting these expressions into
Eq.~\eqref{eq:random_weak_general}, with $x=\cos^2t$, gives
\begin{align}
e(t,\beta)
={}&
1-e^{-c\sin^2t}
\nonumber\\
&+
\beta^2 e^{-c\sin^2t}
\bigg[
t\sin(2t)
\left[
c+c^2\left(1+\cos^2t\right)
\right]
\nonumber\\
&
+
t^2c^2\cos^2t
\bigg]
+
O(\beta^3)
+
O(M^{-1}).
\label{eq:sparse_ER}
\end{align}
The absence of a contribution linear in $\beta$ reflects the vanishing
triangle density of sparse Erd\H{o}s--R\'enyi graphs. The same is true
for four-cycles and overlapping triangles, whose contributions to the
ED density are suppressed as $M^{-1}$. The surviving second-order term
is therefore controlled by the locally tree-like neighborhood: the
Katz interaction generates couplings to vertices at distance two and,
at the same order in the ED, modifies the couplings along the original
edges. Consequently, in the sparse regime the first finite-density correction
to the adjacency dynamics is not associated with short cycles, but with
the branching structure of the local neighborhood.

\paragraph{Dense regime.}
We now consider the dense regime, in which $p=O(1)$ as
$M\to\infty$, so that the mean degree grows as
$\langle d\rangle\simeq pM$. In the same limit,
$\lambda_1\simeq pM$. For a fixed finite $t$, the weak-propagation conditions therefore require $\beta$ to be chosen such that $\beta pM\ll1$ and $t\beta p^2M^2\ll1$. For fixed $x=\cos^2t$, one has $a(x)=1-p+px<1$, and therefore, from Eqs.~\eqref{eq:G_0}--\eqref{eq:G_T2} one obtains
\begin{equation}
    G_0,G_T,G_W,D_2,G_Q,G_{T^2}\longrightarrow0.
\end{equation}
Substituting these expressions into Eq.~\eqref{eq:random_weak_general} gives
\begin{equation}
    e(t,\beta)\longrightarrow1.
\end{equation}
Thus, in the dense regime, the ED density already reaches its maximal value at the adjacency level as a consequence of the extensive connectivity, while the weak-propagation corrections vanish in the
thermodynamic limit.

\section{Discussion and outlook}

We have investigated the entanglement dynamics generated by
Katz-weighted Ising interactions on graphs. A gapped fermionic mediator
provides a microscopic realization of this interaction, whose effective
couplings are determined by the Katz kernel of the underlying graph.
Within the resulting Ising model, the Entanglement Distance can be
evaluated exactly, with the matrix elements of the Katz kernel directly
controlling the graph-dependent entangling phases.

The exact dynamics admits two complementary structural descriptions.
The distance-shell representation organizes the interaction according
to the length and multiplicity of walks connecting different vertices,
whereas the spectral representation resolves the same dynamics into
collective adjacency modes. This distinction becomes particularly
transparent across the two propagation regimes. For
$\beta\ll\beta_c=1/\lambda_1$, the first corrections to the adjacency
interaction can be organized in a finite-time expansion in $\beta$:
triangles enter at first order, while four-cycles, the local degree
structure, and overlapping triangles appear at second order. In the
opposite limit $\beta\to\beta_c^{-}$, the Katz kernel becomes dominated
by the principal adjacency mode, so that the localization properties
of its eigenvector determine the spatial distribution of the
entanglement.

For Erd\H{o}s--R\'enyi graphs, the weak-propagation expansion can be
averaged analytically over the ensemble. In the sparse regime
$p=c/M$, the linear correction vanishes with the triangle density,
while the surviving second-order contribution is controlled by the
locally tree-like branching structure. In the dense regime,
$p=O(1)$, the extensive connectivity already makes the adjacency
contribution dominant in the thermodynamic limit, while the
weak-propagation corrections are suppressed within the controlled
scaling regime.

A natural extension of the present work is to use the dependence of the
entanglement dynamics on the graph structure as a tool for controlled
state preparation. Instead of considering the network as a fixed input,
one may ask which graph geometries and mediator parameters generate a
desired spatial distribution of entanglement or selected multipartite
states. The special-time structures found for highly symmetric graphs
provide simple examples of this possibility. It would also be
interesting to determine how robust these features remain in the
presence of imperfections, such as disorder in the couplings,
inhomogeneities of the mediator, or decoherence. This would clarify to
what extent the walk-based and spectral mechanisms identified here can
be exploited for entanglement generation and distribution in realistic
quantum networks.

\begin{acknowledgments}
We acknowledge the support of the Research Support Plan 2022 – Call for applications for funding allocation to research projects curiosity-driven (F CUR) – Project "Entanglement Protection of Qubits’ Dynamics in a Cavity" – EPQDC and the support from the Italian National Group of Mathematical Physics (GNFM-INdAM). R. F. would like to acknowledge INFN Pisa for the financial support for this activity.
\end{acknowledgments}

\nocite{*}

\bibliography{references}

@article{PhysRevLett.86.910,
  title = {Persistent Entanglement in Arrays of Interacting Particles},
  author = {Briegel, Hans J. and Raussendorf, Robert},
  journal = {Phys. Rev. Lett.},
  volume = {86},
  issue = {5},
  pages = {910--913},
  numpages = {0},
  year = {2001},
  month = {Jan},
  publisher = {American Physical Society},
  doi = {10.1103/PhysRevLett.86.910},
  url = {https://link.aps.org/doi/10.1103/PhysRevLett.86.910}
}

@article{PhysRevLett.86.5188,
  title = {A One-Way Quantum Computer},
  author = {Raussendorf, Robert and Briegel, Hans J.},
  journal = {Phys. Rev. Lett.},
  volume = {86},
  issue = {22},
  pages = {5188--5191},
  numpages = {0},
  year = {2001},
  month = {May},
  publisher = {American Physical Society},
  doi = {10.1103/PhysRevLett.86.5188},
  url = {https://link.aps.org/doi/10.1103/PhysRevLett.86.5188}
}

@article{PhysRevA.69.062311,
  title = {Multiparty entanglement in graph states},
  author = {Hein, M. and Eisert, J. and Briegel, H. J.},
  journal = {Phys. Rev. A},
  volume = {69},
  issue = {6},
  pages = {062311},
  numpages = {20},
  year = {2004},
  month = {Jun},
  publisher = {American Physical Society},
  doi = {10.1103/PhysRevA.69.062311},
  url = {https://link.aps.org/doi/10.1103/PhysRevA.69.062311}
}

@article{PhysRevLett.97.150504,
  title = {Universal Resources for Measurement-Based Quantum Computation},
  author = {Van den Nest, Maarten and Miyake, Akimasa and D\"ur, Wolfgang and Briegel, Hans J.},
  journal = {Phys. Rev. Lett.},
  volume = {97},
  issue = {15},
  pages = {150504},
  numpages = {4},
  year = {2006},
  month = {Oct},
  publisher = {American Physical Society},
  doi = {10.1103/PhysRevLett.97.150504},
  url = {https://link.aps.org/doi/10.1103/PhysRevLett.97.150504}
}

@article{Hartmann_2007,
doi = {10.1088/0953-4075/40/9/S01},
url = {https://doi.org/10.1088/0953-4075/40/9/S01},
year = {2007},
month = {apr},
publisher = {},
volume = {40},
number = {9},
pages = {S1},
author = {Hartmann, L and Calsamiglia, J and Dür, W and Briegel, H J},
title = {Weighted graph states and applications to spin chains, lattices and gases},
journal = {Journal of Physics B: Atomic, Molecular and Optical Physics}
}

@article{Anders_2007,
doi = {10.1088/1367-2630/9/10/361},
url = {https://doi.org/10.1088/1367-2630/9/10/361},
year = {2007},
month = {oct},
publisher = {},
volume = {9},
number = {10},
pages = {361},
author = {Anders, Simon and Briegel, Hans J and Dür, Wolfgang},
title = {A variational method based on weighted graph states},
journal = {New Journal of Physics}
}

@article{Perseguers_2013,
doi = {10.1088/0034-4885/76/9/096001},
url = {https://doi.org/10.1088/0034-4885/76/9/096001},
year = {2013},
month = {sep},
publisher = {IOP Publishing},
volume = {76},
number = {9},
pages = {096001},
author = {Perseguers, S and Lapeyre, G J and Cavalcanti, D and Lewenstein, M and Acín, A},
title = {Distribution of entanglement in large-scale quantum networks},
journal = {Reports on Progress in Physics}
}

@article{Katz_1953, title={A New Status Index Derived from Sociometric Analysis}, volume={18}, DOI={10.1007/BF02289026}, number={1}, journal={Psychometrika}, author={Katz, Leo}, year={1953}, pages={39–43}}

@article{PhysRevE.77.036111,
  title = {Communicability in complex networks},
  author = {Estrada, Ernesto and Hatano, Naomichi},
  journal = {Phys. Rev. E},
  volume = {77},
  issue = {3},
  pages = {036111},
  numpages = {12},
  year = {2008},
  month = {Mar},
  publisher = {American Physical Society},
  doi = {10.1103/PhysRevE.77.036111},
  url = {https://link.aps.org/doi/10.1103/PhysRevE.77.036111}
}

@article{10.1093/comnet/cnt007,
    author = {Benzi, Michele and Klymko, Christine},
    title = {Total communicability as a centrality measure},
    journal = {Journal of Complex Networks},
    volume = {1},
    number = {2},
    pages = {124-149},
    year = {2013},
    month = {12},
    issn = {2051-1310},
    doi = {10.1093/comnet/cnt007},
    url = {https://doi.org/10.1093/comnet/cnt007},
    eprint = {https://academic.oup.com/comnet/article-pdf/1/2/124/7093013/cnt007.pdf},
}

@article{PhysRevLett.103.240503,
  title = {Entanglement Percolation in Quantum Complex Networks},
  author = {Cuquet, Mart\'{\i} and Calsamiglia, John},
  journal = {Phys. Rev. Lett.},
  volume = {103},
  issue = {24},
  pages = {240503},
  numpages = {4},
  year = {2009},
  month = {Dec},
  publisher = {American Physical Society},
  doi = {10.1103/PhysRevLett.103.240503},
  url = {https://link.aps.org/doi/10.1103/PhysRevLett.103.240503}
}

@article{Collins_2013,
doi = {10.1088/1751-8113/46/30/305302},
url = {https://doi.org/10.1088/1751-8113/46/30/305302},
year = {2013},
month = {jul},
publisher = {IOP Publishing},
volume = {46},
number = {30},
pages = {305302},
author = {Collins, Benoît and Nechita, Ion and Życzkowski, Karol},
title = {Area law for random graph states},
journal = {Journal of Physics A: Mathematical and Theoretical}
}

@article{PhysRevA.89.052335,
  title = {Randomized graph states and their entanglement properties},
  author = {Wu, Jun-Yi and Rossi, Matteo and Kampermann, Hermann and Severini, Simone and Kwek, Leong Chuan and Macchiavello, Chiara and Bru\ss{}, Dagmar},
  journal = {Phys. Rev. A},
  volume = {89},
  issue = {5},
  pages = {052335},
  numpages = {15},
  year = {2014},
  month = {May},
  publisher = {American Physical Society},
  doi = {10.1103/PhysRevA.89.052335},
  url = {https://link.aps.org/doi/10.1103/PhysRevA.89.052335}
}

@article{im/1109190962,
author = {Fan Chung and Linyuan Lu and Van Vu},
title = {{The Spectra of Random Graphs with Given Expected Degrees}},
volume = {1},
journal = {Internet Mathematics},
number = {3},
publisher = {A K Peters, Ltd.},
pages = {257 -- 275},
year = {2003},
}

@article{52xz-3hpc,
  title = {Random Regular Graph States Are Complex at Almost Any Depth},
  author = {Ghosh, Soumik and Hangleiter, Dominik and Helsen, Jonas},
  journal = {PRX Quantum},
  volume = {6},
  issue = {4},
  pages = {040344},
  numpages = {40},
  year = {2025},
  month = {Nov},
  publisher = {American Physical Society},
  doi = {10.1103/52xz-3hpc},
  url = {https://link.aps.org/doi/10.1103/52xz-3hpc}
}

@ARTICLE{Wallnofer2019-zs,
  title    = "Multipartite state generation in quantum networks with optimal
              scaling",
  author   = "Walln{\"o}fer, J and Pirker, A and Zwerger, M and D{\"u}r, W",
  journal  = "Scientific Reports",
  volume   =  9,
  number   =  1,
  pages    = "314",
  month    =  jan,
  year     =  2019
}

@ARTICLE{Hahn2019-kw,
  title    = "Quantum network routing and local complementation",
  author   = "Hahn, F and Pappa, A and Eisert, J",
  journal  = "npj Quantum Information",
  volume   =  5,
  number   =  1,
  pages    = "76",
  month    =  sep,
  year     =  2019
}

@article{PhysRevA.108.062614,
  title = {Multiparty entanglement routing in quantum networks},
  author = {Mannalath, Vaisakh and Pathak, Anirban},
  journal = {Phys. Rev. A},
  volume = {108},
  issue = {6},
  pages = {062614},
  numpages = {15},
  year = {2023},
  month = {Dec},
  publisher = {American Physical Society},
  doi = {10.1103/PhysRevA.108.062614},
  url = {https://link.aps.org/doi/10.1103/PhysRevA.108.062614}
}

@article{cocchiarella_entanglement_2020,
	title = {Entanglement distance for arbitrary \${M}\$-qudit hybrid systems},
	volume = {101},
	issn = {2469-9926, 2469-9934},
	url = {http://arxiv.org/abs/2003.05771},
	doi = {10.1103/PhysRevA.101.042129},
	number = {4},
	urldate = {2021-07-05},
	journal = {Phys. Rev. A},
	author = {Cocchiarella, Denise and Scali, Stefano and Ribisi, Salvatore and Nardi, Bianca and Bel-Hadj-Aissa, Ghofrane and Franzosi, Roberto},
	month = apr,
	year = {2020},
	pages = {042129},
}

@article{vafafard_multipartite_2022,
	title = {Multipartite stationary entanglement generation in the presence of dipole-dipole interaction in an optical cavity},
	volume = {105},
	issn = {2469-9926, 2469-9934},
	url = {https://link.aps.org/doi/10.1103/PhysRevA.105.052439},
	doi = {10.1103/PhysRevA.105.052439},
	number = {5},
	urldate = {2023-08-09},
	journal = {Phys. Rev. A},
	author = {Vafafard, Azar and Nourmandipour, Alireza and Franzosi, Roberto},
	month = may,
	year = {2022},
	pages = {052439},
}

@article{nourmandipour_entanglement_2021,
	title = {Entanglement protection of classically driven qubits in a lossy cavity},
	volume = {11},
	issn = {2045-2322},
	url = {https://doi.org/10.1038/s41598-021-95623-1},
	doi = {10.1038/s41598-021-95623-1},
	number = {1},
	journal = {Scientific Reports},
	author = {Nourmandipour, Alireza and Vafafard, Azar and Mortezapour, Ali and Franzosi, Roberto},
	month = aug,
	year = {2021},
	pages = {16259},
}

@article{vesperini_entanglement_2023,
	title = {Entanglement and quantum correlation measures for quantum multipartite mixed states},
	volume = {13},
	issn = {2045-2322},
	url = {https://doi.org/10.1038/s41598-023-29438-7},
	doi = {10.1038/s41598-023-29438-7},
	number = {1},
	journal = {Scientific Reports},
	author = {Vesperini, Arthur and Bel-Hadj-Aissa, Ghofrane and Franzosi, Roberto},
	month = feb,
	year = {2023},
	pages = {2852},
}

@article{vesperini_correlations_2023,
	title = {Correlations and projective measurements in maximally entangled multipartite states},
	volume = {457},
	issn = {0003-4916},
	url = {https://www.sciencedirect.com/science/article/pii/S0003491623001926},
	doi = {https://doi.org/10.1016/j.aop.2023.169406},
	journal = {Annals of Physics},
	author = {Vesperini, Arthur},
	year = {2023},
	pages = {169406},
}

@article{Vesperini_2023,
   title={Entanglement and quantum correlation measures for quantum multipartite mixed states},
   volume={13},
   ISSN={2045-2322},
   url={http://dx.doi.org/10.1038/s41598-023-29438-7},
   DOI={10.1038/s41598-023-29438-7},
   number={1},
   journal={Scientific Reports},
   publisher={Springer Science and Business Media LLC},
   author={Vesperini, Arthur and Bel-Hadj-Aissa, Ghofrane and Franzosi, Roberto},
   year={2023},
   month=feb }

@article{Vesperini_2024,
   title={Unveiling the geometric meaning of quantum entanglement: Discrete and continuous variable systems},
   volume={19},
   ISSN={2095-0470},
   url={http://dx.doi.org/10.1007/s11467-024-1403-x},
   DOI={10.1007/s11467-024-1403-x},
   number={5},
   journal={Frontiers of Physics},
   publisher={Springer Science and Business Media LLC},
   author={Vesperini, Arthur and Bel-Hadj-Aissa, Ghofrane and Capra, Lorenzo and Franzosi, Roberto},
   year={2024},
   month=apr }

@Article{e28030299,
AUTHOR = {De Simone, Lucio and Capra, Lorenzo and Vesperini, Arthur and Rossi, Leonardo and Di Cairano, Loris and Franzosi, Roberto},
TITLE = {Geometric Aspects of Entanglement},
JOURNAL = {Entropy},
VOLUME = {28},
YEAR = {2026},
NUMBER = {3},
ARTICLE-NUMBER = {299},
URL = {https://www.mdpi.com/1099-4300/28/3/299},
PubMedID = {41899951},
ISSN = {1099-4300},
DOI = {10.3390/e28030299}
}

@article{DeSimone_2025,
doi = {10.1088/1751-8121/ae0bcb},
url = {https://doi.org/10.1088/1751-8121/ae0bcb},
year = {2025},
month = {oct},
publisher = {IOP Publishing},
volume = {58},
number = {41},
pages = {415302},
author = {De Simone, Lucio and Franzosi, Roberto},
title = {Entanglement in directed graph states},
journal = {Journal of Physics A: Mathematical and Theoretical}
}

@article{https://doi.org/10.1002/qute.202500514,
author = {De Simone, Lucio and Franzosi, Roberto},
title = {Entanglement in Quantum Systems Based on Directed Graphs},
journal = {Advanced Quantum Technologies},
volume = {9},
number = {1},
pages = {e00514},
doi = {https://doi.org/10.1002/qute.202500514},
url = {https://advanced.onlinelibrary.wiley.com/doi/abs/10.1002/qute.202500514},
year = {2026}
}

@article{PhysRevLett.110.216803,
  title = {Majorana-Klein Hybridization in Topological Superconductor Junctions},
  author = {B\'eri, B.},
  journal = {Phys. Rev. Lett.},
  volume = {110},
  issue = {21},
  pages = {216803},
  numpages = {5},
  year = {2013},
  month = {May},
  publisher = {American Physical Society},
  doi = {10.1103/PhysRevLett.110.216803},
  url = {https://link.aps.org/doi/10.1103/PhysRevLett.110.216803}
}

@misc{hein2006entanglementgraphstatesapplications,
      title={Entanglement in Graph States and its Applications}, 
      author={M. Hein and W. Dür and J. Eisert and R. Raussendorf and M. Van den Nest and H. -J. Briegel},
      year={2006},
      eprint={quant-ph/0602096},
      archivePrefix={arXiv},
      primaryClass={quant-ph},
      url={https://arxiv.org/abs/quant-ph/0602096}, 
}

@article{Frucht_1949, title={Graphs of Degree Three with a Given Abstract Group}, volume={1}, DOI={10.4153/CJM-1949-033-6}, number={4}, journal={Canadian Journal of Mathematics}, author={Frucht, Robert}, year={1949}, pages={365–378}}

@article{ESTRADA201289,
title = {The physics of communicability in complex networks},
journal = {Physics Reports},
volume = {514},
number = {3},
pages = {89-119},
year = {2012},
note = {The Physics of Communicability in Complex Networks},
issn = {0370-1573},
doi = {https://doi.org/10.1016/j.physrep.2012.01.006},
url = {https://www.sciencedirect.com/science/article/pii/S0370157312000154},
author = {Ernesto Estrada and Naomichi Hatano and Michele Benzi}
}

\end{document}